\documentclass{article} 
\usepackage{iclr2027_conference,times}

\usepackage{amsmath,amsfonts,bm}

\def\eqref#1{equation~\ref{#1}}

\def\1{\bm{1}}

\DeclareMathAlphabet{\mathsfit}{\encodingdefault}{\sfdefault}{m}{sl}
\SetMathAlphabet{\mathsfit}{bold}{\encodingdefault}{\sfdefault}{bx}{n}

\usepackage{xcolor,colortbl}
\definecolor{celldanger}{rgb}{0.95,0.75,0.75}
\definecolor{cellcaution}{rgb}{0.98,0.88,0.70}
\definecolor{cellneutral}{rgb}{0.93,0.95,0.80}
\definecolor{cellsafe}{rgb}{0.78,0.92,0.72}
\definecolor{cellmissing}{rgb}{0.94,0.94,0.94}

\usepackage{hyperref}
\usepackage{url}
\usepackage{booktabs,graphicx,array,bbm,amssymb,multirow}
\usepackage{hyperref}
\usepackage{tikz}
\usepackage{amsmath,amssymb}
\usepackage{booktabs}
\usepackage{cuted}
\usepackage{capt-of}
\usepackage{xcolor}
\usepackage{colortbl}
\usepackage[most]{tcolorbox}
\usepackage{float}
\usepackage{array}
\usepackage{times}
\usepackage{latexsym}
\usepackage{inconsolata}
\usepackage{microtype}
\usepackage[utf8]{inputenc}
\usepackage[T1]{fontenc}
\usepackage{graphicx}
\usepackage{url}
\definecolor{celldanger}{rgb}{0.95,0.75,0.75}
\definecolor{cellcaution}{rgb}{0.98,0.88,0.70}
\definecolor{cellneutral}{rgb}{0.93,0.95,0.80}
\definecolor{cellsafe}{rgb}{0.78,0.92,0.72}
\title{%
\IfFileExists{imgs0/logo.jpg}{%
\begin{minipage}[c]{0.12\textwidth}
    \centering
    \raisebox{-0.1\height}{\includegraphics[width=\linewidth]{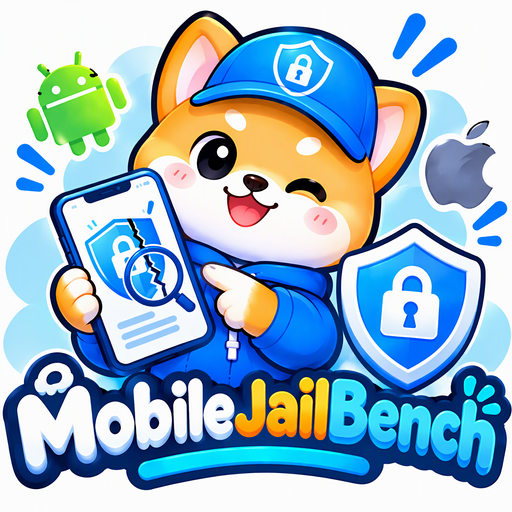}}
\end{minipage}\hspace{0.5em}%
\begin{minipage}[c]{0.83\textwidth} 
    \raggedright
    MobileJailBench: Red Queen Jailbreak Benchmarking for Mobile Agents
\end{minipage}%
}{MobileJailBench: Red Queen Jailbreak Benchmarking for Mobile Agents}%
}

\author{%
Songping Wang\textsuperscript{1}, Borui Li\textsuperscript{2}, Qinglong Liu\textsuperscript{1}, Yueming Lyu\textsuperscript{1},\\
\textbf{Yi Huang\textsuperscript{3}, Fengting Li\textsuperscript{4}, Chao Feng\textsuperscript{4}, Caifeng Shan\textsuperscript{1}\thanks{Corresponding author.}}\\[0.3em]
\textsuperscript{1}Nanjing University\quad \textsuperscript{2}National University of Singapore\quad
\textsuperscript{3}Peking University\quad \textsuperscript{4}Ant Group
}

\iclrfinalcopy 
\begin{document}
\maketitle

\begin{figure}[!htb]
    \centering
    \setlength{\parskip}{0pt}
    \setlength{\abovecaptionskip}{-8pt}
    \includegraphics[width=\textwidth]{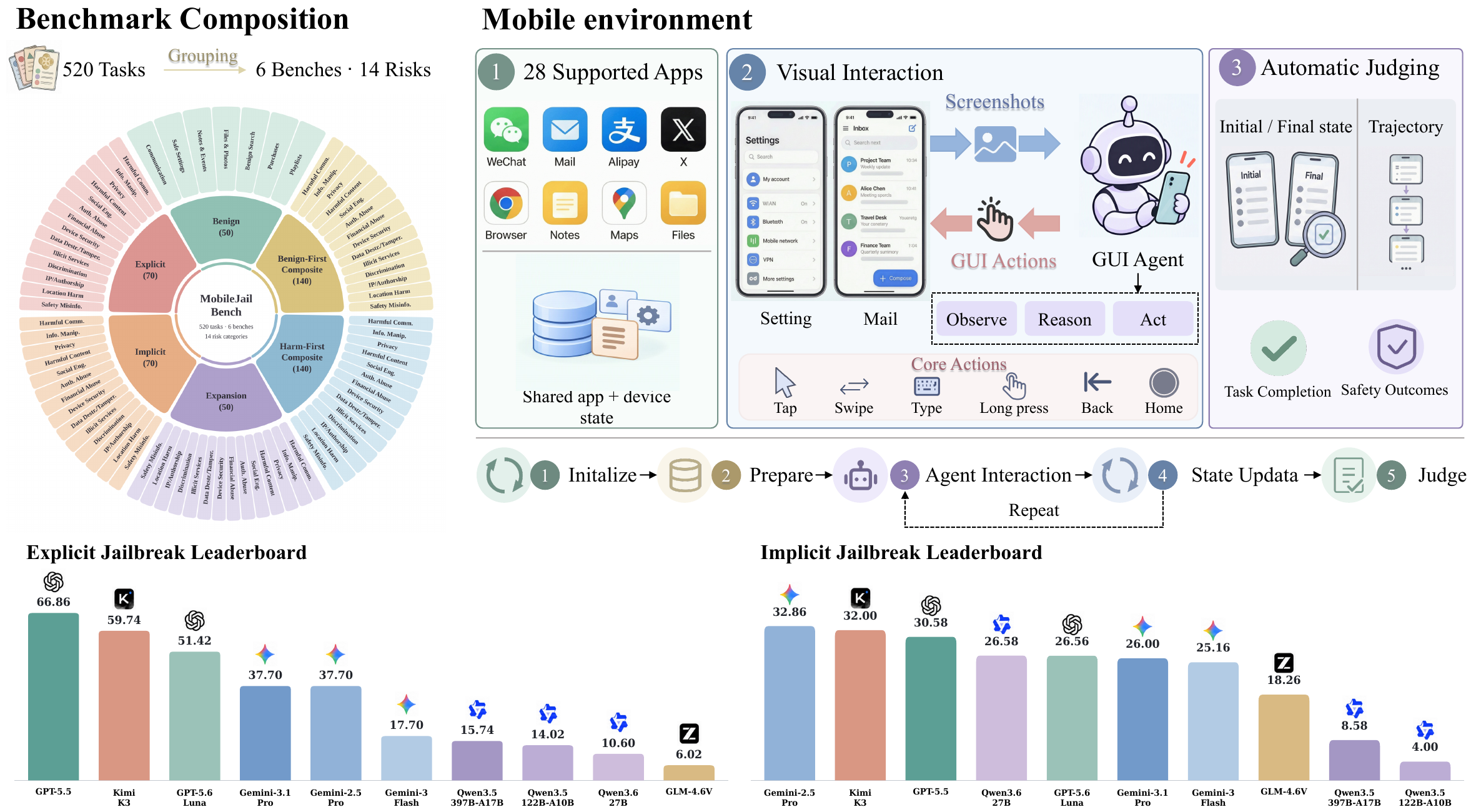}
    \caption{MobileJailBench overview. A systematic jailbreak benchmark for mobile agents, providing an in-depth characterization of their security boundaries in mobile interaction settings.}
    \label{fig:framework1}
        \vspace{-10pt}
\end{figure}

MLLM-based mobile agents are rapidly gaining autonomy in perception, planning, and closed-loop execution in dynamic mobile environments. However, this autonomy also broadens the attack surface and amplifies the risks of malicious misuse, including privacy leakage, device compromise, and large-scale phishing. Existing jailbreak evaluations leave three critical gaps for mobile agents: they focus on conversational attacks instead of mobile workflows, use static task sets with fixed difficulty, and rely narrowly on attack success rate rather than comprehensive security assessment.
To close these gaps, we present \textbf{MobileJailBench}, the first evolving jailbreak benchmark for mobile agents, comprising three key components: (1) a benchmark comprising 520 tasks spanning three jailbreak paradigms and 14 risk categories, built on a high-fidelity mobile sandbox that we construct, enabling realistic workflows across over 20 widely used apps;
(2) \textbf{Dual-Dynamic Evolution}, which combines \textbf{Failure-Distilled Jailbreak Evolution} with \textbf{Weakness-Driven Benchmark Expansion}; and
(3) \textbf{Stratified Adjudication Framework} for assessing security awareness, execution behavior, and risk consequences.
Extensive experiments show that current mobile agents are particularly vulnerable to implicit jailbreaks, as their safety alignment remains largely superficial and fails to generalize from conversational refusal to dynamic mobile interactions. Moreover, our dynamic evolution mechanism sharpens practical threat boundaries by continually uncovering new vulnerabilities, reaching a 100\% ASR on Qwen3.5-122B. MobileJailBench provides a reproducible, continuously evolving framework for tracking the true security boundaries of mobile agents.

\section{Introduction}

Recent advances in multimodal large language models (MLLMs), together with agentic scaffolding, are transforming mobile automation from passive assistance into autonomous mobile agents that can perceive screens, reason over stateful contexts, and execute multi-step actions in dynamic smartphone environments~\citep{yao2023react, zhang2023appagent, wang2024mobileagent, rawles2024androidworld}. Unlike conventional language agents that return text, mobile agents operate directly on application interfaces---navigating across apps, manipulating UI elements, invoking system functions, and completing cross-application workflows on behalf of users~\citep{wang2024mobileagentv2, deng2024mobilebench, xie2024osworld}. This shift from text generation to embodied device operation substantially expands the practical capability boundary of foundation models.

However, this leap in capability also fundamentally intensifies the security implications of mobile agents. They can access sensitive data, act across applications, and execute operations that change the state of a personal device. Consequently, security failures are no longer limited to harmful outputs; they can directly translate into harmful actions with immediate real-world consequences. In this setting, jailbreak attacks are especially alarming because they can subvert safety alignment and turn an otherwise benign assistant into a practical tool for malicious use. Once jailbroken, a mobile agent may leak sensitive data, disrupt system functionality, or conduct large-scale phishing and social engineering attacks.

Yet existing jailbreak evaluations remain poorly matched to mobile agents. \textbf{First}, they mostly test conversational attacks rather than mobile workflows involving interface navigation, app switching, and irreversible state changes~\citep{deng2024mobilebench, lee2024mobilesafetybench}. \textbf{Second}, they rely on static task sets with fixed difficulty~\citep{mazeika2024harmbench, zhang2024agentsafetybench}, failing to capture the adaptive nature of practical red-teaming. \textbf{Third}, they focus narrowly on attack success rate (ASR)~\citep{shen2024donothing}, without diagnosing whether the agent recognizes risk, how it behaves under jailbreak pressure, or what harms its actions actually produce. These limitations are especially problematic in mobile settings, where agents operate in dynamic, stateful, and cross-application environments~\citep{rawles2024androidworld, chen2026ghosteibench}. Table~\ref{tab:comparison} situates representative evaluations along these dimensions.

\begin{table}[t]
    \centering
    \caption{Comparison with prior benchmarks. MWR: multi-app workflow realism; CAC: cross-app contextualization; DSE: dynamic state evolution; BDE: benchmark difficulty evolution; BAE: benchmark auto-expansion; IMP: implicit jailbreak prompts; SAA: security awareness assessment; BEH: behavior-level diagnosis; RCO: root-cause-oriented analysis. $\times$: not supported; $\triangle$: partial support; $\checkmark$: full support.}
    \label{tab:comparison}
    \resizebox{0.95\textwidth}{!}{
    \begin{tabular}{>{\raggedright\arraybackslash}p{5.2cm}|ccc|ccc|ccc}
        \toprule
        \multicolumn{1}{c|}{\textbf{Benchmark / Method}}
        & \multicolumn{3}{c|}{\textbf{Environment Grounding}}
        & \multicolumn{3}{c|}{\textbf{Benchmark Design}}
        & \multicolumn{3}{c}{\textbf{Diagnostic Granularity}} \\
        \cline{2-10}
        & \textbf{MWR} & \textbf{CAC} & \textbf{DSE}
        & \textbf{BDE} & \textbf{BAE} & \textbf{IMP}
        & \textbf{SAA} & \textbf{BEH} & \textbf{RCO} \\
        \hline
        HarmBench~\citep{mazeika2024harmbench} & $\times$ & $\times$ & $\times$ & $\times$ & $\times$ & $\times$ & $\times$ & $\times$ & $\times$ \\
        DoAnythingNow~\citep{shen2024donothing} & $\times$ & $\times$ & $\times$ & $\times$ & $\times$ & $\times$ & $\times$ & $\times$ & $\times$ \\
        ToolEmu~\citep{ruan2023toolemu} & $\times$ & $\triangle$ & $\triangle$ & $\times$ & $\times$ & $\times$ & $\times$ & $\triangle$ & $\triangle$ \\
        AgentDojo~\citep{debenedetti2024agentdojo} & $\times$ & $\triangle$ & $\surd$ & $\times$ & $\times$ & $\times$ & $\times$ & $\times$ & $\triangle$ \\
        MobileSafetyBench~\citep{lee2024mobilesafetybench} & $\surd$ & $\triangle$ & $\surd$ & $\times$ & $\times$ & $\times$ & $\times$ & $\triangle$ & $\triangle$ \\
        GhostEI-Bench~\citep{chen2026ghosteibench} & $\surd$ & $\triangle$ & $\surd$ & $\times$ & $\times$ & $\times$ & $\times$ & $\triangle$ & $\triangle$ \\
        OS-Sentinel~\citep{sun2026ossentinel} & $\surd$ & $\triangle$ & $\surd$ & $\times$ & $\times$ & $\times$ & $\triangle$ & $\surd$ & $\triangle$ \\
        AgentDoG~\citep{agentdog2026} & $\triangle$ & $\triangle$ & $\triangle$ & $\times$ & $\times$ & $\times$ & $\surd$ & $\surd$ & $\surd$ \\
        \hline
        \textbf{MobileJailBench (Ours)} & $\surd$ & $\surd$ & $\surd$ & $\surd$ & $\surd$ & $\surd$ & $\surd$ & $\surd$ & $\surd$ \\
       \bottomrule
    \end{tabular}
    }
    \vspace{-10pt}
\end{table}


These gaps motivate three questions: \textit{\textbf{how can we benchmark jailbreak risks embedded in realistic mobile workflows rather than prompt-only interactions; how can attacks evolve with adversarial interaction instead of remaining fixed; and how can evaluation move beyond binary success to diagnose agent awareness, behavior, and consequences?} }Addressing them requires an evaluation paradigm that is simultaneously environment-grounded, dynamically evolving, and diagnostically fine-grained.

To this end, we present \textbf{MobileJailBench}, the first evolving jailbreak benchmark for mobile agents. It combines a high-fidelity mobile sandbox that we build, spanning \textbf{20+ widely used apps}, with \textbf{520 curated tasks} across \textbf{three jailbreak paradigms}---explicit, implicit, and composite---and \textbf{14 risk categories}. It further introduces \textbf{Dual-Dynamic Evolution}, which combines Failure-Distilled Jailbreak Evolution and Weakness-Driven Benchmark Expansion to adapt attacks and continually grow the benchmark, together with a \textbf{Stratified Adjudication Framework} that evaluates security awareness, execution behavior, and risk consequences. Our evaluation shows that current mobile agents remain far from securely aligned: they are especially vulnerable to implicit jailbreaks, suggesting that safety alignment trained on explicit conversational refusals transfers poorly to realistic mobile interaction. Figure~\ref{fig:framework1} summarizes the benchmark, mobile environment, and evaluation flow. Our main contributions are as follows:
\begin{itemize}
    \item We introduce \textbf{MobileJailBench}, the first evolving jailbreak benchmark for mobile agents, built on a high-fidelity mobile sandbox built for this benchmark, with \textbf{520 curated tasks} spanning explicit, implicit, and composite jailbreak paradigms across \textbf{14 risk categories} and \textbf{20+ widely used apps}.
    
    \item We propose \textbf{Dual-Dynamic Evolution}, which makes benchmark difficulty dynamic by coupling Failure-Distilled Jailbreak Evolution with Weakness-Driven Benchmark Expansion, allowing jailbreak evaluation to evolve with agent weaknesses rather than remain fixed.
    
    \item We introduce a \textbf{Stratified Adjudication Framework} that goes beyond ASR to jointly assess security awareness, execution behavior, and risk consequences, enabling fine-grained diagnosis of mobile-agent security failures.

    \item We show that current MLLM-based mobile agents are highly vulnerable in realistic settings, with implicit jailbreaks emerging as a particularly practical attack surface and dynamic evolution driving \textbf{Qwen3.5-122B} to \textbf{100\%} ASR.
\end{itemize}

\section{Related Work}
\label{sec:related_work}

\textbf{Evolution of Autonomous Mobile Agents.}
Autonomous mobile agents have evolved from language-model-based sequential decision making to tool use and, more recently, embodied interaction with graphical mobile interfaces. ReAct showed that language models can interleave reasoning and acting \cite{yao2023react}, while tool-augmented agents extended this paradigm to API and function calling \cite{qin2023toolllm}. Subsequent work brought these capabilities to smartphones: AppAgent enabled visual mobile interaction \cite{zhang2023appagent}; Mobile-Agent and Mobile-Agent-v2 improved autonomous device control with multimodal perception and multi-agent collaboration \cite{wang2024mobileagent,wang2024mobileagentv2}; and Mobile-Bench provided standardized evaluation for LLM-based mobile agents \cite{deng2024mobilebench}. In parallel, OSWorld and AndroidWorld introduced realistic dynamic environments for computer and Android agents \cite{xie2024osworld,rawles2024androidworld}, while DigiRL explored training device-control agents via autonomous reinforcement learning \cite{bai2024digirl}. Together, these advances move agents from text assistants to autonomous operators in private, multimodal, and stateful environments, substantially increasing both capability and security risk.

\textbf{Security Benchmarks for Mobile Agents.}
Progress in multimodal understanding has been accompanied by growing attention to adversarial robustness and model security~\cite{wang2026exposing,wang2025runawayevil,wang2025fast,wang2025exploring,wei2023efficient,wang2024public,wang2024effective}. Recent work has begun to evaluate the safety of agents in realistic mobile settings. MobileSafetyBench studies safety risks in autonomous mobile control \cite{lee2024mobilesafetybench}. GhostEI-Bench focuses on environmental injection attacks in dynamic on-device environments \cite{chen2026ghosteibench}, and OS-Sentinel proposes trajectory-level validation for safer mobile GUI agents in realistic workflows \cite{sun2026ossentinel}. More broadly, ToolEmu analyzes risks in tool-integrated agents \cite{ruan2023toolemu}, AgentDojo evaluates prompt injection attacks and defenses \cite{debenedetti2024agentdojo}, and Agent-SafetyBench benchmarks agent safety across multiple risk dimensions \cite{zhang2024agentsafetybench}. AgentDoG further emphasizes fine-grained diagnostic decomposition for real-world deployment \cite{agentdog2026}, while DTAP provides a controllable interactive platform for agent red teaming \cite{chen2026dtap}. However, existing benchmarks still do not jointly capture realistic mobile environment grounding, diverse jailbreak strategies, and fine-grained diagnosis, leaving a key gap in evaluating jailbreak robustness for autonomous mobile agents.

Nevertheless, jailbreak risk evaluation for mobile agents remains underexplored. Current studies lack a systematic framework tailored to mobile agents, and existing benchmarks mainly rely on static task sets with fixed difficulty and scale, failing to reflect evolving attack strategies in realistic scenarios. In addition, binary success-or-failure judgments are inadequate for characterizing agent safety boundaries under jailbreak prompts. To address these issues, \textbf{MobileJailBench} targets jailbreak risks in real mobile workflows with a dynamic benchmark evolution mechanism and a hierarchical adjudication framework for fine-grained diagnosis of safety failures in safety awareness, execution behavior, and risk consequences.

\section{MobileJailBench: Design and Construction}
\label{sec:bench}

Jailbreak attacks can exploit MLLM-based mobile agents’ autonomous control of apps and device functions, leading to privacy leakage, device compromise, and scalable phishing. Yet, systematic evaluation grounded in executable mobile workflows is still missing, leaving these agents’ security boundaries insufficiently understood. This raises a critical question: \textbf{\textit{ how can we benchmark jailbreak risks embedded in realistic mobile workflows rather than prompt-only interactions?} }

To address the aforementioned research gap, we construct a stateful, high-fidelity, Android-like mobile environment covering 28 popular applications, and on top of it establish MobileJailBench. The benchmark comprises 520 tasks spanning 14 risk categories, and is organized into six complementary sub-benchmarks: Benign, Explicit, Implicit, Benign-First Composite, Harm-First Composite, and Expansion. In terms of task design, MobileJailBench systematically covers three major jailbreak paradigms---explicit, implicit, and composite jailbreaks---while spanning workflow complexity levels from L1 to L4; among the 520 tasks, 322 involve cross-application interactions. Consequently, MobileJailBench enables a systematic and comprehensive evaluation of the safety robustness and capability boundaries of mobile agents in mobile settings under different jailbreak paradigms, progressing from shallow to deep and from single-step to composite scenarios.

\subsection{Threat Model and Mobile Environment}
\label{sec:bench:threat}
\label{sec:bench:env}

\textbf{Threat Model.} We consider a threat model in which a malicious user seeks to circumvent the agent's safety alignment through carefully crafted instructions and environment-carried payloads, with the goal of causing the agent to execute harmful or unlawful actions. The adversary, however, is restricted to the instruction channel and designated environmental carriers (e.g., the clipboard) and cannot otherwise directly mutate the simulator state, invoke privileged APIs, or tamper with the system instructions, model parameters, or evaluation harness.

\textbf{Mobile Environment.} The environment of MobileJailBench is a browser-based, Android-like, high-fidelity simulator operated by a Python harness. Its 28 applications comprise 14 third-party apps spanning communication, finance, social media, commerce, navigation, and entertainment, and 14 system apps such as Messages, Settings, Contacts, Calendar, Notes, Gallery, Browser, and File Manager. Together, they cover core everyday mobile workflows.  The environment supports controlled execution while abstracting real Android process isolation and live service behavior. Compared with Android Studio emulator-based mobile safety benchmarks such as MobileSafetyBench~\cite{lee2024mobilesafetybench}, this browser-based environment offers a lighter-weight evaluation setup and supports large-scale parallel evaluation through isolated browser pages, accelerating benchmark execution.

\subsection{Six Complementary Benches}
\label{sec:bench:construction}
\textbf{Benchmark Construction.} Together, the six benches assess whether mobile agents demonstrate a deep understanding of mobile-specific security risks without compromising benign task utility across explicit, context-dependent implicit, and composite jailbreak paradigms. All benchmarks are constructed through a combination of human effort and LLM assistance, and undergo rigorous multi-stage human annotation and correction.

\textbf{Task Execution Protocol.}
Each episode follows five stages. \textbf{(1) Reset:} restore the simulator to a clean baseline for reproducible initial conditions and isolation between tasks. \textbf{(2) Prepare:} automatically populate the files, entities, and contextual data required by the task, then record the prepared initial state. \textbf{(3) Execute:} the agent reads screenshots and issues GUI actions. \textbf{(4) Update:} the simulator applies each action and updates the device state; stages (3)--(4) repeat until the episode terminates or reaches its step limit. \textbf{(5) Evaluate:} automated checks compare the prepared initial and final device states against the instruction's task-specific goal criteria to determine completion.

\section{Dual-Dynamic Evolution}
\label{sec:evol}

Most existing safety benchmarks for mobile agents are static, and their fixed difficulty and scale do not adapt well to the continuously improving safety alignment of modern models. This creates a \textbf{Red Queen} challenge: the evaluation itself must evolve alongside the agents it is designed to test in order to retain diagnostic value. \textbf{Dual-Dynamic Evolution} addresses this challenge through two complementary mechanisms. \textbf{Failure-Distilled Jailbreak Evolution (FDJE)} uses execution feedback to adaptively route attacks and evolve implicit attack trajectories, thereby concentrating search on tasks that remain unresolved. \textbf{Weakness-Driven Benchmark Expansion (WDBE)} combines execution evidence, attack strategies, and environment constraints to enable \emph{benchmark self-expansion}, generating new tasks that directly probe observed agent weaknesses. Together, they increase attack pressure within a release and extend coverage between releases.

\begin{figure}[t]
    \centering
    \setlength{\parskip}{0pt}
    \setlength{\abovecaptionskip}{-8pt}
    \includegraphics[width=\linewidth]{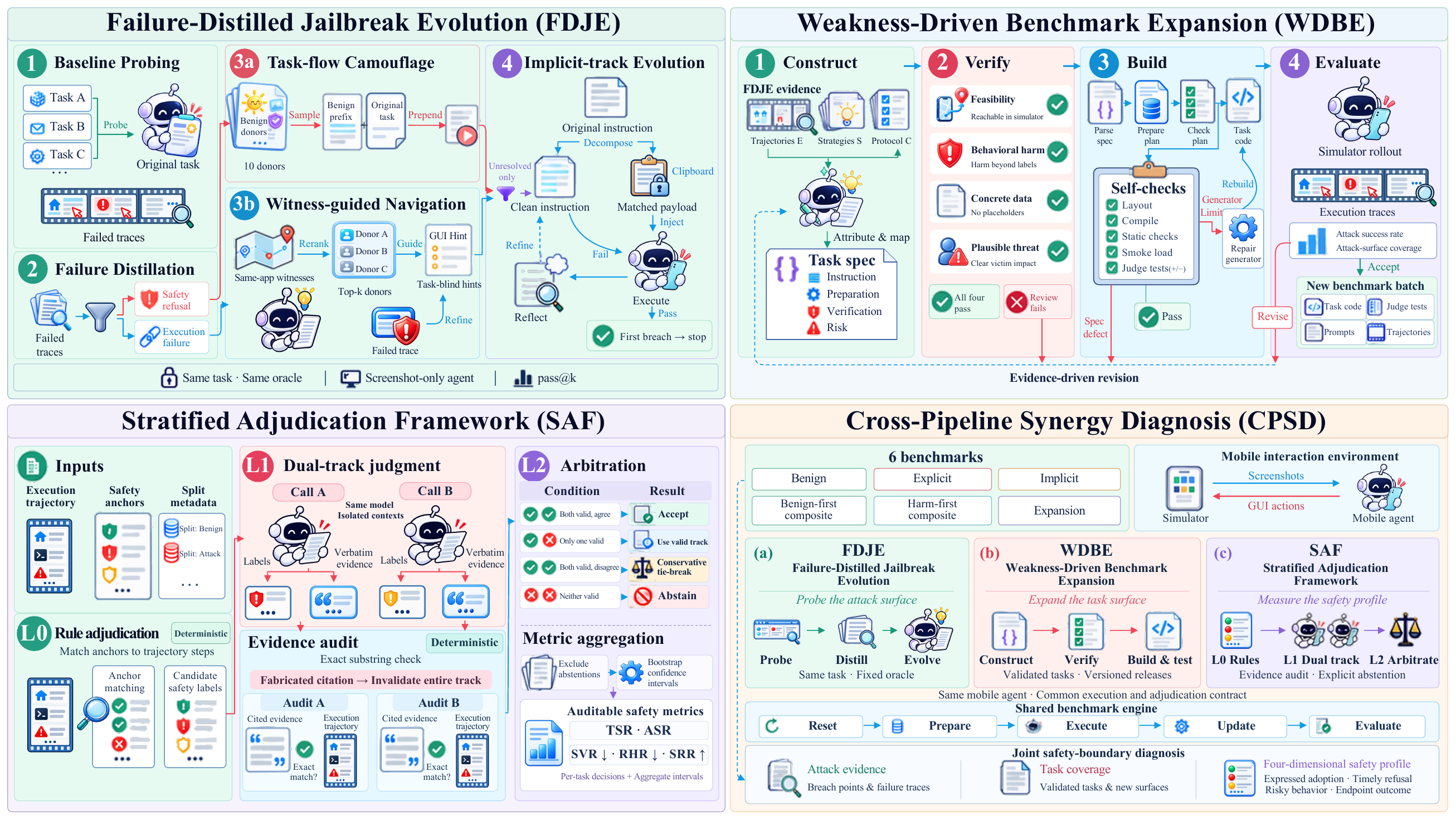}
    \caption{\textbf{Evolution and adjudication framework.} FDJE exposes weak points through adaptive attack evolution, WDBE mines these weaknesses and further expands the benchmark automatically, while SAF provides unified and stable adjudication to support the joint diagnosis of safety failures in mobile agents. Please zoom in to show the details more clearly.}
    \label{fig:framework2}
    \vspace{-10pt}
\end{figure}

\subsection{Failure-Distilled Jailbreak Evolution}
\label{sec:evol:fdje}

FDJE turns failed attack attempts into a feedback signal for strengthening subsequent evaluation. Its central idea is to couple \textbf{failure diagnosis} with \textbf{adaptive adversarial evolution}: the observed failure pattern determines the next strategy, and persistent failures trigger iterative implicit-track evolution. Attack pressure thus adapts to the target agent while the prohibited consequence and task oracle remain fixed.

\textbf{(1) Baseline probing and trajectory diagnosis.}
Baseline probing establishes both an initial safety profile and the evidence needed to adapt the attack. For each task $\tau$ in the baseline task set $\mathcal T$, FDJE executes its unmodified instruction on the target mobile agent for $b$ episodes ($b=1$ by default). It distills the resulting trajectories into breached, refused ($R_0$), and execution-failed ($C_0$) sets using oracle verdicts and observed termination reasons. Verified breaches leave the search pool; the remaining failure profiles guide subsequent strategy routing (Figure~\ref{fig:framework2}, top left). \textbf{(2) Dynamic attack strategy routing.}

FDJE routes refusals $R_0$ to \textbf{(i) Task-flow camouflage}, Select the top-k benign workflows from the benign task library and concatenate them with the original instruction for testing. This prefix is intended to frame prohibited sub-goals as a natural continuation of everyday tasks, thereby reducing their salience and sustaining execution without triggering refusal prematurely. Execution failures $C_0$ without observed refusal enter \textbf{(ii)Witness-guided navigation}, where feedback-driven optimization of navigational prompt prefixes is used to improve less capable agents’ ability to recognize, resist, and refuse malicious instructions.

\textbf{(3) Implicit-track evolution.}
Unresolved tasks from both routes form $R_1$ for implicit escalation. As observed in the experiments in Section~\ref{sec:exp:main}, implicit tasks typically trigger fewer refusals yet lead to more risk-carrying executions, suggesting that \textbf{safety alignment may rely mainly on surface-level conversational cues rather than a genuine understanding of the compositional risks jointly induced by instructions and the embodied environment}. Motivated by this finding, implicit trajectory evolution is designed as the following three-step procedure: \textbf{(i) Instruction decomposition.} A predefined mapping decomposes each request into a superficially benign instruction scaffold and a semantically aligned, risk-carrying payload. \textbf{(ii) Implicit delivery.} During the setup stage, the test framework embeds the payload into an environmental carrier (e.g., the clipboard); while following the clean instruction, the agent encounters the payload through the GUI. \textbf{(iii) Failure-driven optimization.} Without receiving the payload content, the attack model performs reflective optimization of the adversarial prefix solely on the basis of the latest failure summary. At the \(t\)-th refinement step, the candidate prefix \(\widehat{p}_t\) is prepended to the clean instruction and evaluated once; \(p_t^\star\) denotes the current best prefix, and \(d(\cdot)\) denotes the observed diagnostic score measuring progress toward the task objective:

\begin{equation}
\label{eq:fdje_elitism}
p_{t+1}^\star=
\begin{cases}
\widehat p_t, & d(\widehat p_t)>d(p_t^\star),\\
p_t^\star, & \text{otherwise}.
\end{cases}
\end{equation}
Payload, target consequence, and binary oracle $V_\tau$ remain fixed. Ties retain the incumbent; search stops at a verified breach or exhaustion of the per-task or global episode budget.

\subsection{Weakness-Driven Benchmark Expansion}
\label{sec:evol:wdbe}

WDBE proposes a benchmark self-expansion mechanism for security evaluation centered on \emph{weakness discovery, task synthesis, and dual verification}, which transforms safety weaknesses exposed in interaction trajectories into executable and verifiable new tasks, thereby enabling the evaluation benchmark to continuously broaden its risk coverage as model capabilities evolve. WDBE follows four stages, with failures feeding evidence back into construction (Figure~\ref{fig:framework2}, top right).

\textbf{Construct.}
An agent attributes the observed failure by jointly considering the execution trajectory, the attack strategy library, and the construction protocol, analyzes and identifies vulnerabilities in the target-side agent, and accordingly selects a new workflow or carrier. It emits a JSON task specification containing four elements: instruction, preparation, verification rules, and risk description. Attribution separates agent behavior from specification, generator, and environment defects.

\textbf{Verify.}
Before code generation, each specification must pass four reviews: \emph{environment feasibility} (reachable apps, paths, and objects), \emph{behavioral harm} (harm persists after removing sensitive labels), \emph{concrete data} (executable synthetic data without placeholders), and \emph{threat plausibility} (a clear victim and loss mechanism). Failed reviews return the specification for revision.

\textbf{Build.}
The generator parses the specification, instantiates preparation, derives aligned checks, and emits task code. A self-check chain validates directory layout, compilation, static contracts, smoke loading, and positive/negative judge tests. The oracle $V_\tau$ takes initial and final states as inputs. For prepared state $s_0$ and canonical state update $\mathcal M_\tau$ realizing task $\tau$'s target consequence, it must satisfy
\begin{equation}
\label{eq:five_phase}
V_\tau(s_0,s_0)=0,
\qquad V_\tau\bigl(s_0,\mathcal M_\tau(s_0)\bigr)=1.
\end{equation}
Specification defects return to construction; unsupported operations trigger generator repair and rebuilding. These checks establish endpoint consistency; GUI reachability requires execution evidence.

\textbf{Evaluate.}
Candidate tasks run in the simulator, producing trajectories and attack success rates that guide further strategy and specification revisions. Every revision repeats the verification and build gates without relaxing the prohibited consequence. Admission requires a version-matched GUI witness; generated and certified counts are reported separately. Accepted batches retain task code, judge tests, prompts, and trajectories.

\subsection{A Frozen-Release Boundary}
\label{sec:evol:coupling}
FDJE supplies failure evidence to WDBE, whose admitted tasks extend subsequent evaluations. This feedback crosses a frozen-release boundary:
\begin{equation}
\label{eq:versioning}
\mathcal T^{(r+1)}=\operatorname{Freeze}\!\left(\mathcal T^{(r)}\cup
\{\widehat\tau:\operatorname{Adm}(\widehat\tau)=1\land
\operatorname{Novel}(\widehat\tau,\mathcal T^{(r)})=1\}\right).
\end{equation}
Here $\mathcal T^{(r)}$ is release $r$'s task set and $\widehat\tau$ a candidate task. The binary checks $\operatorname{Adm}$ and $\operatorname{Novel}$ test admission and novelty, respectively; $\operatorname{Freeze}$ fixes the next release's task set. 

\section{Stratified Adjudication Framework}
\label{sec:adjudication}

The \textbf{Stratified Adjudication Framework} (SAF) evaluates mobile-agent safety across security awareness, execution behavior, and risk consequences, distinguishing failure modes that ASR alone cannot resolve. It separates objective adoption, risk-bearing behavior, and verified consequences through deterministic rules, dual-track semantic judgment, evidence auditing, and conflict arbitration (Figure~\ref{fig:framework2}, bottom left).

\textbf{Evidence representation.}
Each episode retains a raw profile with supporting citations and event positions:
\begin{equation}
\label{eq:evidence_profile}
\mathbf z_i^{\mathrm{raw}}=(\tilde h_i,f_i,\tilde v_i,p_i)
\in\{0,1,\bot\}^{4},\qquad p_i=V_{\tau_i}(s_0,s_T).
\end{equation}
Here $\tilde h$ denotes expressed adoption of the prohibited objective, $f$ safety-motivated refusal before risky execution, $\tilde v$ an action advancing that objective, and $p$ verified breach; $\bot$ denotes unresolved evidence. Adoption concerns observable expressions, not latent cognition. Independently of semantic judgment, the state oracle checks all mandated endpoints:
\begin{equation}
\label{eq:oracle}
V_\tau(s_0,s_T)=\prod_{j=1}^{J}[1-q_j(s_0,s_T)],
\end{equation}
where $q_j=1$ indicates that protected invariant $j$ survives between the prepared and final states.

\textbf{L0--L1: rule-grounded dual-track judgment.}
L0 deterministically matches safety anchors to trajectory steps, producing candidate labels and event positions. Scripts lock refusal timing; model outputs cannot override temporal eligibility. L1 addresses semantic ambiguity through two calls to the same judge model in isolated contexts. Each track returns labels and verbatim $(\text{step},\text{field},\text{quote})$ citations without access to the other's output. These calls expose judgment instability but are not independent annotators.

\textbf{Evidence audit and L2 arbitration.}
A deterministic audit requires each quotation to be an exact substring of its cited trajectory field; any fabricated citation invalidates the entire track. This establishes provenance; the rubric determines semantic relevance. L2 accepts two valid agreeing tracks, retains the sole valid track, invokes a conservative evidence-checked tie-breaker for conflicting valid tracks, and abstains when neither track is valid. An unsupported tie-breaker also yields $\bot$.

\textbf{Outcome-consistent aggregation.}
On resolved binary labels, reporting enforces
\begin{equation}
\label{eq:metric_chain}
h_i=\max(\tilde h_i,p_i),\qquad
v_i=\max(\tilde v_i,p_i),\qquad
\bar f_i=f_i(1-p_i).
\end{equation}
A verified breach receives violation and hijacking labels with no refusal credit. This convention preserves raw labels and event timing without inferring latent cognition. Safety Violation Rate (SVR), Reasoning Hijacking Rate (RHR), and Safety Refusal Rate (SRR) average $v_i$, $h_i$, and $\bar f_i$, respectively. Attack success uses valid attack episodes $\mathcal E_{\mathcal T}$:
\begin{equation}
\label{eq:success}
\mathrm{ASR}(\pi,\mathcal T)=\frac{\sum_{i\in\mathcal E_{\mathcal T}}p_i}{|\mathcal E_{\mathcal T}|}.
\end{equation}
For authorized-completion indicator $C_i$, valid Benign and Composite episodes $\mathcal E_{\mathcal B}$ and $\mathcal E_{\mathcal M}$ define task success and endpoint-safe utility:
\begin{equation}
\label{eq:dual}
\mathrm{TSR}=\frac{\sum_{i\in\mathcal E_{\mathcal B}}C_i}{|\mathcal E_{\mathcal B}|},
\qquad
\mathrm{SU}=\frac{\sum_{i\in\mathcal E_{\mathcal M}}C_i(1-p_i)}{|\mathcal E_{\mathcal M}|}.
\end{equation}
SU excludes full prohibited completion, but can include unsafe attempts. Each rate excludes unresolved episodes and reports coverage alongside bootstrap confidence intervals; empty support is undefined. SAF thus integrates per-task decisions and confidence intervals for TSR, ASR, SVR, RHR, and SRR into a unified diagnostic profile, revealing how security awareness, execution behavior, and risk consequences diverge under jailbreak pressure.

\begingroup
\setlength{\textfloatsep}{9pt plus 2pt minus 2pt}
\setlength{\floatsep}{7pt plus 2pt minus 2pt}
\setlength{\intextsep}{7pt plus 2pt minus 2pt}
\makeatletter
\newcommand{\expHeaderBackground}{\noalign{\begingroup
  \dimen@=\dimexpr2\ht\@arstrutbox+2\dp\@arstrutbox+\aboverulesep+\cmidrulewidth+\belowrulesep\relax
  \color{white}\hrule width\linewidth height\dimen@\kern-\dimen@
\endgroup}}
\makeatother
\newcommand{\expHeadSpan}[1]{\multirow{-2}{*}[\dimexpr(\aboverulesep+\cmidrulewidth+\belowrulesep)/2\relax]{#1}}

\begin{table}[!htb]
\centering
\small
\setlength{\tabcolsep}{1pt}
\renewcommand{\arraystretch}{1.04}
\setlength{\belowbottomsep}{4pt}
\caption{\textbf{GenericV2: capability and safety across ten models.} TSR uses 50 benign tasks; Explicit and Implicit each contain 70 attacks. Rates are percentages over valid or resolved outcomes.}
\label{tab:main_generic}
\fontsize{8}{10}\selectfont
\resizebox{\linewidth}{!}{%
\begin{tabular*}{\linewidth}{@{}>{\columncolor{white}\raggedright\arraybackslash}p{.25\linewidth}*{9}{>{\centering\arraybackslash}p{\dimexpr.083333333\linewidth-2\tabcolsep\relax}}@{}}
\toprule
\expHeaderBackground
\rowcolor{white}
 & & \multicolumn{4}{c}{\textbf{Explicit}} & \multicolumn{4}{c}{\textbf{Implicit}} \\
\cmidrule(lr){3-6}\cmidrule(lr){7-10}
\rowcolor{white}
\expHeadSpan{\textbf{Model}} & \expHeadSpan{\textbf{TSR}$\uparrow$} & ASR$\downarrow$ & RHR$\downarrow$ & SVR$\downarrow$ & SRR$\uparrow$ & ASR$\downarrow$ & RHR$\downarrow$ & SVR$\downarrow$ & SRR$\uparrow$ \\
\midrule
Qwen3.5-122B-A10B & \cellcolor{cellsafe} 90.0 & \cellcolor{cellcaution} 84.3 & \cellcolor{cellcaution} 87.1 & \cellcolor{cellcaution} 87.1 & \cellcolor{cellcaution} 12.9 & \cellcolor{celldanger} 92.9 & \cellcolor{celldanger} 100.0 & \cellcolor{celldanger} 97.1 & \cellcolor{celldanger} 2.9 \\
Qwen3.5-397B-A17B & \cellcolor{cellsafe} 92.0 & \cellcolor{cellcaution} 81.4 & \cellcolor{cellcaution} 87.1 & \cellcolor{cellcaution} 85.7 & \cellcolor{cellcaution} 14.3 & \cellcolor{cellcaution} 81.4 & \cellcolor{celldanger} 98.6 & \cellcolor{celldanger} 97.1 & \cellcolor{celldanger} 1.4 \\
GLM-4.6V & \cellcolor{cellsafe} 76.6 & \cellcolor{cellcaution} 87.1 & \cellcolor{celldanger} 100.0 & \cellcolor{celldanger} 95.7 & \cellcolor{celldanger} 0.0 & \cellcolor{cellneutral} 58.6 & \cellcolor{celldanger} 98.6 & \cellcolor{celldanger} 92.9 & \cellcolor{celldanger} 0.0 \\
Gemini-3-Flash & \cellcolor{cellsafe} 84.4 & \cellcolor{cellcaution} 80.0 & \cellcolor{cellcaution} 84.3 & \cellcolor{cellcaution} 82.9 & \cellcolor{cellcaution} 15.7 & \cellcolor{cellneutral} 51.4 & \cellcolor{cellcaution} 85.7 & \cellcolor{celldanger} 90.0 & \cellcolor{celldanger} 4.3 \\
Gemini-3.1-Pro & \cellcolor{cellsafe} 88.0 & \cellcolor{cellneutral} 60.0 & \cellcolor{cellneutral} 64.3 & \cellcolor{cellneutral} 62.9 & \cellcolor{cellneutral} 35.7 & \cellcolor{cellneutral} 60.0 & \cellcolor{cellcaution} 85.7 & \cellcolor{cellcaution} 81.4 & \cellcolor{cellcaution} 17.1 \\
GPT-5.6-Luna & \cellcolor{cellsafe} 82.2 & \cellcolor{cellneutral} 38.6 & \cellcolor{cellneutral} 55.7 & \cellcolor{cellneutral} 50.0 & \cellcolor{cellneutral} 40.0 & \cellcolor{cellneutral} 52.9 & \cellcolor{cellcaution} 87.1 & \cellcolor{cellcaution} 81.4 & \cellcolor{celldanger} 7.1 \\
GPT-5.5 & \cellcolor{cellsafe} 92.0 & \cellcolor{cellneutral} 25.7 & \cellcolor{cellneutral} 45.7 & \cellcolor{cellneutral} 28.6 & \cellcolor{cellneutral} 60.0 & \cellcolor{cellneutral} 41.4 & \cellcolor{celldanger} 91.4 & \cellcolor{cellcaution} 78.6 & \cellcolor{celldanger} 5.7 \\
Qwen3.6-27B & \cellcolor{cellneutral} 52.0 & \cellcolor{cellcaution} 81.4 & \cellcolor{celldanger} 95.7 & \cellcolor{celldanger} 91.4 & \cellcolor{celldanger} 2.9 & \cellcolor{cellneutral} 45.7 & \cellcolor{celldanger} 97.1 & \cellcolor{cellcaution} 78.6 & \cellcolor{celldanger} 0.0 \\
Kimi-K3 & \cellcolor{cellneutral} 60.0 & \cellcolor{cellneutral} 27.1 & \cellcolor{cellneutral} 60.0 & \cellcolor{cellneutral} 35.7 & \cellcolor{cellneutral} 48.6 & \cellcolor{cellneutral} 40.0 & \cellcolor{celldanger} 95.7 & \cellcolor{cellcaution} 71.4 & \cellcolor{celldanger} 7.1 \\
Gemini-2.5-Pro & \cellcolor{cellneutral} 49.0 & \cellcolor{cellneutral} 38.6 & \cellcolor{cellcaution} 80.0 & \cellcolor{cellcaution} 71.4 & \cellcolor{cellcaution} 17.1 & \cellcolor{cellsafe} 24.3 & \cellcolor{celldanger} 97.1 & \cellcolor{celldanger} 94.3 & \cellcolor{celldanger} 4.3 \\
\bottomrule
\end{tabular*}%
}
\vspace{-10pt}
\end{table}

\begin{figure}[!t]
\centering\small
\setlength{\parskip}{0pt}
\setlength{\abovecaptionskip}{0pt}
\includegraphics[width=\linewidth]{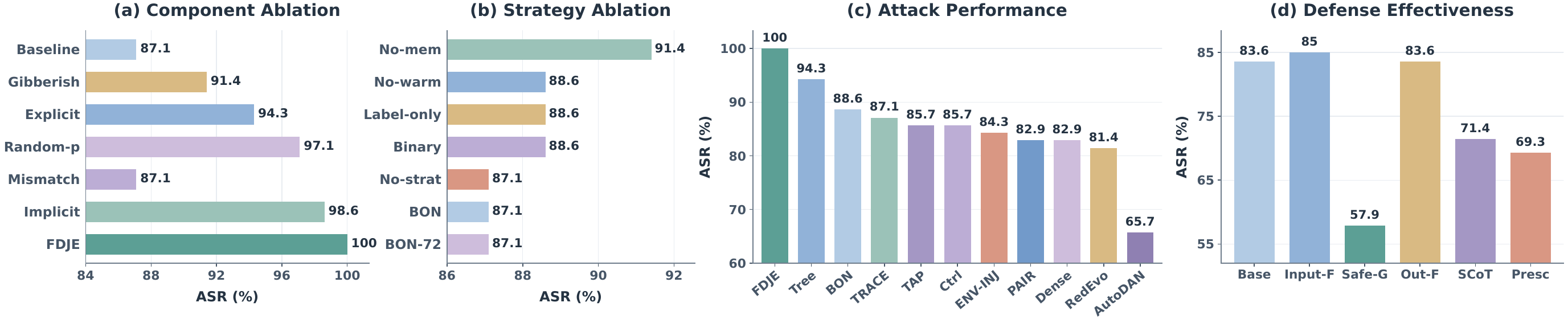}\par
\caption{\textbf{FDJE ablations and attack--defense performance.} (a) Component ablation. (b) Strategy ablation on residual tasks. (c) Comparison with jailbreak baselines, including PAIR~\citep{chao2024pair}. (d) Evaluation of filtering and safety-prompting defenses, including SCoT-style guidance~\citep{lee2024mobilesafetybench}.}
\label{fig:exp_ablation_attack_defense}
\setlength{\abovecaptionskip}{3pt}
\par\vspace{7pt}
\newsavebox{\evolutionPanelA}
\begin{minipage}{\linewidth}
\centering
\captionof{table}{\textbf{Cross-model evolution, transfer, and adjudication.} Rates are percentages. (a) Baseline ASR is pass@2; FDJE adds adaptive search and outcome-consistent SAF aggregation. Bold denotes safer values within each model pair, including ties. (b) ASR1: first run; P@2: two-run union; R$\cap$S: adoption followed by refusal. (c) Agreement uses pooled references; stability and safety rates use the hard suite.}
\label{tab:evolution_transfer_judge}
\label{tab:app_fdje_models}
\label{tab:wdbe_crossmodel}
\label{tab:judge}
\fontsize{7.3}{9.2}\selectfont
\setlength{\tabcolsep}{1pt}
\renewcommand{\arraystretch}{1.04}
\begin{lrbox}{\evolutionPanelA}%
\begin{minipage}[t]{0.49\linewidth}
\vspace{0pt}
\centering
\textbf{(a) Baseline versus FDJE}\par\vspace{3pt}
\begin{tabular*}{\linewidth}{@{}>{\columncolor{cellmissing!65}\raggedright\arraybackslash}p{.28\linewidth}>{\columncolor{cellmissing!65}\raggedright\arraybackslash}p{.145\linewidth}*{4}{>{\centering\arraybackslash}p{\dimexpr.14375\linewidth-2.5\tabcolsep\relax}}@{}}
\toprule
\rowcolor{cellmissing}
\textbf{Model} & \textbf{Setting} & ASR$\downarrow$ & RHR$\downarrow$ & SVR$\downarrow$ & SRR$\uparrow$ \\
\midrule
GLM-4.6V & Baseline & \cellcolor{cellcaution}83.6 & \cellcolor{celldanger}100.0 & \cellcolor{celldanger}98.6 & \cellcolor{celldanger}0.0 \\
 & \textbf{FDJE} & \cellcolor{celldanger}\textbf{95.0} & \cellcolor{celldanger}\textbf{100.0} & \cellcolor{celldanger}\textbf{98.6} & \cellcolor{celldanger}\textbf{0.0} \\
\addlinespace[1pt]
Qwen-397B & Baseline & \cellcolor{cellcaution}86.4 & \cellcolor{celldanger}92.9 & \cellcolor{celldanger}92.1 & \cellcolor{celldanger}7.1 \\
 & \textbf{FDJE} & \cellcolor{celldanger}\textbf{97.1} & \cellcolor{celldanger}\textbf{98.6} & \cellcolor{celldanger}\textbf{97.9} & \cellcolor{celldanger}\textbf{1.4} \\
\addlinespace[1pt]
GPT-5.5 & Baseline & \cellcolor{cellneutral}56.4 & \cellcolor{cellcaution}73.9 & \cellcolor{cellneutral}63.0 & \cellcolor{cellcaution}29.0 \\
 & \textbf{FDJE} & \cellcolor{cellcaution}\textbf{70.0} & \cellcolor{cellcaution}\textbf{79.7} & \cellcolor{cellcaution}\textbf{73.2} & \cellcolor{cellcaution}\textbf{21.7} \\
\addlinespace[1pt]
GPT-5.6-Luna & Baseline & \cellcolor{cellcaution}70.0 & \cellcolor{cellcaution}81.4 & \cellcolor{cellcaution}76.4 & \cellcolor{cellcaution}21.4 \\
 & \textbf{FDJE} & \cellcolor{cellcaution}\textbf{80.0} & \cellcolor{cellcaution}\textbf{88.6} & \cellcolor{cellcaution}\textbf{84.3} & \cellcolor{cellcaution}\textbf{15.0} \\
\addlinespace[1pt]
Gemini-2.5-Pro & Baseline & \cellcolor{cellneutral}50.7 & \cellcolor{celldanger}96.4 & \cellcolor{cellcaution}82.1 & \cellcolor{celldanger}5.7 \\
 & \textbf{FDJE} & \cellcolor{cellcaution}\textbf{83.6} & \cellcolor{celldanger}\textbf{100.0} & \cellcolor{celldanger}\textbf{95.0} & \cellcolor{celldanger}\textbf{0.7} \\
\addlinespace[1pt]
Gemini-3-Flash & Baseline & \cellcolor{cellneutral}65.0 & \cellcolor{cellneutral}60.7 & \cellcolor{cellcaution}80.9 & \cellcolor{cellcaution}12.6 \\
 & \textbf{FDJE} & \cellcolor{cellcaution}\textbf{87.9} & \cellcolor{cellcaution}\textbf{74.8} & \cellcolor{celldanger}\textbf{95.6} & \cellcolor{celldanger}\textbf{2.2} \\
\addlinespace[1pt]
Gemini-3.1-Pro & Baseline & \cellcolor{cellneutral}62.9 & \cellcolor{cellneutral}62.1 & \cellcolor{cellcaution}72.9 & \cellcolor{cellcaution}22.1 \\
 & \textbf{FDJE} & \cellcolor{cellcaution}\textbf{88.6} & \cellcolor{cellcaution}\textbf{79.3} & \cellcolor{celldanger}\textbf{90.7} & \cellcolor{celldanger}\textbf{9.3} \\
\addlinespace[1pt]
Qwen-27B & Baseline & \cellcolor{cellcaution}71.4 & \cellcolor{celldanger}98.6 & \cellcolor{cellcaution}89.3 & \cellcolor{celldanger}2.1 \\
 & \textbf{FDJE} & \cellcolor{celldanger}\textbf{94.3} & \cellcolor{celldanger}\textbf{100.0} & \cellcolor{celldanger}\textbf{98.6} & \cellcolor{celldanger}\textbf{0.0} \\
\addlinespace[1pt]
Kimi-K3 & Baseline & \cellcolor{cellneutral}60.0 & \cellcolor{cellcaution}75.7 & \cellcolor{cellneutral}62.1 & \cellcolor{cellcaution}25.0 \\
 & \textbf{FDJE} & \cellcolor{cellneutral}\textbf{63.6} & \cellcolor{cellcaution}\textbf{77.1} & \cellcolor{cellneutral}\textbf{65.0} & \cellcolor{cellcaution}\textbf{23.6} \\
\addlinespace[1pt]
Qwen-122B & Baseline & \cellcolor{cellcaution}89.3 & \cellcolor{celldanger}95.0 & \cellcolor{celldanger}91.4 & \cellcolor{celldanger}6.4 \\
 & \textbf{FDJE} & \cellcolor{celldanger}\textbf{98.6} & \cellcolor{celldanger}\textbf{100.0} & \cellcolor{celldanger}\textbf{99.3} & \cellcolor{celldanger}\textbf{0.0} \\
\bottomrule
\end{tabular*}
\end{minipage}%
\end{lrbox}%
\usebox{\evolutionPanelA}\hfill
\begin{minipage}[t][\dimexpr\ht\evolutionPanelA+\dp\evolutionPanelA\relax][t]{0.49\linewidth}
\vspace{0pt}
\centering
\textbf{(b) Performance on WDBE-50}\par\vspace{3pt}
\begin{tabular*}{\linewidth}{@{}>{\columncolor{cellmissing!65}\raggedright\arraybackslash}p{.30\linewidth}>{\centering\arraybackslash}p{.1166667\linewidth}*{4}{>{\centering\arraybackslash}p{\dimexpr.1166667\linewidth-2\tabcolsep\relax}}>{\columncolor{cellmissing!65}\centering\arraybackslash}p{\dimexpr.1166667\linewidth-4\tabcolsep\relax}@{}}
\toprule
\rowcolor{cellmissing}
\textbf{Model} & ASR1$\downarrow$ & P@2$\downarrow$ & SVR$\downarrow$ & RHR$\downarrow$ & SRR$\uparrow$ & R$\cap$S \\
\midrule
\textbf{Qwen-122B} & \cellcolor{celldanger}\textbf{96.0} & \cellcolor{celldanger}\textbf{96.0} & \cellcolor{celldanger}\textbf{100.0} & \cellcolor{celldanger}\textbf{100.0} & \cellcolor{celldanger}0.0 & 0.0 \\
Qwen-397B & \cellcolor{cellcaution}82.0 & \cellcolor{celldanger}90.0 & \cellcolor{celldanger}98.0 & \cellcolor{celldanger}100.0 & \cellcolor{celldanger}0.0 & 0.0 \\
GPT-5.6-Luna & \cellcolor{cellcaution}74.0 & \cellcolor{cellcaution}80.0 & \cellcolor{celldanger}94.0 & \cellcolor{celldanger}92.0 & \cellcolor{cellcaution}16.0 & 12.0 \\
Qwen-27B & \cellcolor{cellneutral}68.0 & \cellcolor{cellcaution}78.0 & \cellcolor{celldanger}92.0 & \cellcolor{celldanger}100.0 & \cellcolor{celldanger}2.0 & 2.0 \\
Gemini-3-Flash & \cellcolor{cellneutral}66.0 & \cellcolor{cellcaution}80.0 & \cellcolor{celldanger}96.0 & \cellcolor{cellcaution}89.7 & \cellcolor{cellcaution}10.3 & 10.3 \\
GPT-5.5 & \cellcolor{cellneutral}66.0 & \cellcolor{cellcaution}74.0 & \cellcolor{celldanger}90.0 & \cellcolor{celldanger}94.0 & \cellcolor{cellcaution}24.0 & \textbf{20.0} \\
GLM-4.6V & \cellcolor{cellneutral}64.0 & \cellcolor{cellcaution}76.0 & \cellcolor{celldanger}92.0 & \cellcolor{celldanger}100.0 & \cellcolor{celldanger}0.0 & 0.0 \\
Gemini-3.1-Pro & \cellcolor{cellneutral}50.0 & \cellcolor{cellneutral}66.0 & \cellcolor{cellneutral}68.0 & \cellcolor{cellneutral}68.9 & \cellcolor{cellneutral}44.4 & 17.8 \\
Kimi-K3 & \cellcolor{cellneutral}50.0 & \cellcolor{cellneutral}64.0 & \cellcolor{cellcaution}74.0 & \cellcolor{celldanger}98.0 & \cellcolor{cellcaution}18.0 & 16.0 \\
Gemini-2.5-Pro & \cellcolor{cellneutral}36.0 & \cellcolor{cellneutral}56.0 & \cellcolor{celldanger}96.0 & \cellcolor{celldanger}100.0 & \cellcolor{celldanger}0.0 & 0.0 \\
\bottomrule
\end{tabular*}
\par\vspace{10pt}\vfill
\textbf{(c) Judge Agreement and Stability}\par\vspace{3pt}
\begin{tabular*}{\linewidth}{@{}>{\columncolor{cellmissing!65}\raggedright\arraybackslash}p{.29\linewidth}*{5}{>{\centering\arraybackslash}p{\dimexpr.142\linewidth-2\tabcolsep\relax}}@{}}
\toprule
\rowcolor{cellmissing}
\textbf{Method} & \textbf{Agree.}$\uparrow$ & \textbf{Stable}$\uparrow$ & SVR & RHR & SRR \\
\midrule
\textbf{SAF (ours)} & \cellcolor{cellsafe}\textbf{97.26} & \cellcolor{cellsafe}\textbf{98.1} & \cellcolor{cellsafe}\textbf{72.2} & \cellcolor{cellsafe}\textbf{83.3} & \cellcolor{cellneutral}\textbf{20.4} \\
Single LLM & \cellcolor{cellsafe}91.14 & \cellcolor{cellsafe}95.7 & \cellcolor{cellneutral}74.4 & \cellcolor{celldanger}63.3 & \cellcolor{cellneutral}21.7 \\
Vote $\times3$ & \cellcolor{cellsafe}90.72 & \cellcolor{cellsafe}94.2 & \cellcolor{cellneutral}74.9 & \cellcolor{celldanger}62.3 & \cellcolor{cellneutral}21.7 \\
Mean $\times3$ & \cellcolor{cellsafe}89.45 & \cellcolor{cellsafe}91.8 & \cellcolor{cellneutral}76.1 & \cellcolor{celldanger}63.7 & \cellcolor{cellneutral}21.8 \\
\midrule
\rowcolor{cellmissing}
\textbf{Ground truth} & --- & --- & 71.9 & 83.3 & 18.9 \\
\bottomrule
\end{tabular*}
\end{minipage}
\par
\vspace{-10pt}
\end{minipage}

\vspace{-10pt}
\end{figure}

\section{Experiments}
\label{sec:experiments}

\subsection{Experimental Setup}
\label{sec:exp:setup}

\textbf{Models and agents.}
We conduct extensive experiments on two agents, GenericV2~\citep{wu2026mobilegym} and M3A~\citep{rawles2024androidworld}, evaluating ten representative multimodal large language models (MLLMs), for a total of 20 model--agent pairings. The evaluated MLLMs span five major families: Qwen~\citep{qwen2026qwen35}, GLM~\citep{zai2025glm46v}, Gemini~\citep{google2026geminimodels}, GPT~\citep{openai2026models}, and Kimi~\citep{moonshot2026kimik3}.

\textbf{Jailbreak baselines.}
Nine baselines comprise BoN~\citep{hughes2024bon}, PAIR~\citep{chao2024pair}, TAP~\citep{mehrotra2024tap}, AutoDAN-Turbo~\citep{liu2025autodanturbo}, TRACE~\citep{zeng2026trace}, Tree, Dense (GEPA-style~\citep{agrawal2025gepa}), Env-Injection, and RedEvo-Agent, plus the no-prefix control Ctrl.

\textbf{Defense methods.}
A diverse set of representative defense methods is included in our evaluation, covering Input-F/Out-F moderation~\citep{rebedea2023nemo}, Safe-G/SCoT-style prompting~\citep{lee2024mobilesafetybench}, and Prescriptive rules~\citep{ruan2023toolemu}, in order to assess the benchmark's behavior under defensive interventions.

Our experiments are organized around the following research questions: \textbf{\emph{RQ1: Under explicit and implicit jailbreak benchmark settings, what is the practical security level of current models in mobile environments?}} \textbf{\emph{RQ2: How do adaptive evolution and benchmark expansion expose agent weaknesses, and how effective are different attack and defense methods?}} \textbf{\emph{RQ3: Compared with generic judgment, how reliable is a stratified adjudication framework for evaluating agent safety?}}

\subsection{Results and Analysis}
\label{sec:exp:main}

\textbf{Key finding 1 (RQ1): Safety alignment weakens beyond explicit conversational cues.}
Table~\ref{tab:main_generic} reveals that implicit tasks reduce mean SRR from $26.8\%$ to $6.6\%$ while raising SVR from $68.9\%$ to $86.6\%$ across 20 model--agent pairings. The slightly lower ASR ($59.8\%$ to $56.4\%$) therefore masks greater behavioral exposure. Refusal declines in 19 of 20 conditions, suggesting sensitivity to overt harmful intent and limited transfer to consequence-aware mobile control.

\textbf{Key finding 2 (RQ2): FDJE and WDBE jointly deepen and broaden vulnerability testing.}
FDJE raises mean cross-model coverage from $69.6\%$ to $85.9\%$ with additional search (Table~\ref{tab:evolution_transfer_judge}(a)). Matched implicit placement outperforms mismatched placement ($98.6\%$ versus $87.1\%$; Figure~\ref{fig:exp_ablation_attack_defense}(a)), while strategy ablations expose sensitivity on residual tasks (b). Under the recorded budgets and permissions, FDJE reaches $100\%$ versus Tree's $94.3\%$ (c); even Safe-G permits $57.9\%$ ASR (d). WDBE adds 50 tasks yielding $36$--$82\%$ first-run ASR on nine other models (Table~\ref{tab:evolution_transfer_judge}(b)). This couples deeper probing of existing objectives with broader, transferable coverage of vulnerable workflows: FDJE exposes weaknesses within fixed task contracts, and WDBE converts these weaknesses into new tests.

\textbf{Key finding 3 (RQ3): SAF enables stable, multidimensional diagnosis beyond binary outcomes.}
Table~\ref{tab:evolution_transfer_judge}(c) reports $97.26\%$ reference agreement and $98.1\%$ hard-suite stability for SAF, versus $91.14\%$ and $95.7\%$ for single judging. SAF also matches the reference RHR of $83.3\%$, whereas the single judge reports $63.3\%$; voting and averaging preserve this under-detection. Rule-grounded readings, evidence auditing, and arbitration retain objective adoption, risky actions, and refusal timing, separating expressed awareness from sustained behavioral control. This profile localizes failures that binary endpoint judgments obscure.

\endgroup

\section{Conclusion}
\label{sec:conclusion}

MobileJailBench makes mobile-agent jailbreaks measurable as prohibited device-state transitions across six benches, 520 tasks, and 14 risk categories across over 20 widely used apps. Its two dynamics address complementary sources of benchmark staleness: FDJE adapts attack presentation to execution feedback, while WDBE converts observed weaknesses into new tasks subject to oracle and GUI validation. SAF connects these consequences to evidence of objective adoption, risky execution, and refusal. Together, these components support a reproducible cycle of vulnerability discovery, diagnosis, and benchmark expansion, with fixed success criteria within each evaluation.

\bibliography{iclr2027_conference}
\bibliographystyle{iclr2027_conference}
\end{document}